# Demonstrating 100 Gbps in and out of the public Clouds

Igor Sfiligoi
University of California San Diego
La Jolla CA USA
isfiligoi@sdsc.edu

## ABSTRACT

There is increased awareness and recognition that public Cloud providers do provide capabilities not found elsewhere, with elasticity being a major driver. The value of elastic scaling is however tightly coupled to the capabilities of the networks that connect all involved resources, both in the public Clouds and at the various research institutions. This paper presents results of measurements involving file transfers inside public Cloud providers, fetching data from on-prem resources into public Cloud instances and fetching data from public Cloud storage into on-prem nodes. The networking of the three major Cloud providers, namely Amazon Web Services, Microsoft Azure and the Google Cloud Platform, has been benchmarked. The on-prem nodes were managed by either the Pacific Research Platform or located at the University of Wisconsin – Madison. The observed sustained throughput was of the order of 100 Gbps in all the tests moving data in and out of the public Clouds and throughput reaching into the Tbps range for data movements inside the public Cloud providers themselves. All the tests used HTTP as the transfer protocol.

## 1 Introduction

Scientific computing needs are growing dramatically with time and many communities have to occasionally deal with workloads that exceed the capacity of their local compute resource. At the same time, public Cloud computing has been gaining traction, including funding agencies starting to invest in this sector; examples being NSF's E-CAS and CloudBank awards, and the European Cloud Initiative. Cloud computing, with its promise of elasticity, is the ideal platform for accommodating occasional spikes in computing needs. Other papers showed that the amount of Compute available in public Clouds is quite substantial, with one experiment aggregating 380 fp32 PFLOP32s of GPU-based compute in a single compute pool [1]. The value of elastic scaling is however tightly coupled to the capabilities of the networks that connect all involved resources, both in the public Clouds and at the various research institutions.

An important aspect of any major scientific computing project is data movement. Researchers and their support teams need to understand the basic characteristics of the networks attached to the resources they will use in order to make proper planning decisions regarding the resources to use. In the case of commercial Cloud resources, while the performance and cost of compute instances is relatively well documented and understood, the same cannot be said for network links and data movement at large. To address this deficiency, several network characterization campaigns have been done, transferring data both inside the single public Cloud providers and moving data between Cloud and on-prem resources. The networking of the three major Cloud providers, namely Amazon Web Services (AWS), Microsoft Azure and the Google Cloud Platform (GCP), has been benchmarked. The first round of results was published in [2], while this paper provides the description of the larger scale tests, which happened mostly after that first paper was written.

Section 3 describes the experience of pulling data from on-prem storage into public Cloud instances, while Section 4 describes pulling data from Cloud storage into on-prem compute nodes. In both cases sustained throughput of around 100 Gbps have been observed.

Section 2 provides a description of network measurements taken within single cloud Provider resources. This information is provided mostly as proof that networking inside a single Cloud provider substantially exceeds what is achievable once leaving that domain.

### 1.1 Related work

The author is not aware of any other recent large-scale hybrid-Cloud network characterization paper.

## 2 Networking inside a single Cloud provider

When measuring the throughput of end-to-end network flows, the obtained throughput will be limited by the slowest link in the path. In order to set the upper bound for data flowing in and out for the Clouds, the capabilities of in-Cloud networking must be measured.

In order to measure the performance of in-Cloud networking, a scalable endpoint had to be picked. The Cloud-operated object storage was selected, since all public Cloud providers operate one as a large pool of distributed storage and offer an easy-to-access object storage interface. Moreover, many scientific compute workloads are likely to access Cloud-native object storage, too, so measuring achievable performance against it was important in its own right. The test setup was the same for all of the commercial Cloud providers. We created a set of files, each 1 GB in size, and uploaded them to the object storage in one Cloud region per tested commercial Cloud provider, namely one in AWS, one in Azure and one in GCP. We then provisioned a set of compute instances and started an increasingly large number of concurrent download processes on them, collecting the timing logs to measure the achieved performance. The observed aggregated throughput exceeded 1 Tbps for all the tested public Cloud providers, see Figure 1, and higher throughputs are likely achievable with increased parallelism.

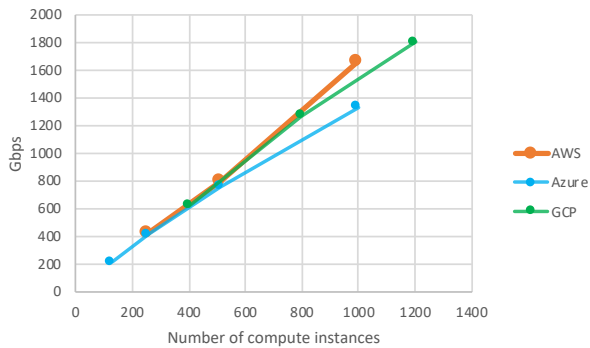

**Figure 1: Peak Throughput observed in a Cloud region while downloading from a local object storage instance, for each of the three tested commercial Cloud providers.**

A similar test was then performed using object storage in one geographical region and compute instances in another region. The number of pair-wise tests is too high to provide a detailed analysis of the results, but throughputs varied between approximately 100 Gbps and 1 Tbps. The peak observed throughput between the major regions is available in Table 1.

*Table 1: Peak Throughput observed between regions of a single public Cloud provider.*

|                | AWS      | Azure    | GCP       |
|----------------|----------|----------|-----------|
| **Asia – US West** | 100 Gbps | 110 Gbps | 940 Gbps  |
| **US East – West** | 440 Gbps | 190 Gbps | 1060 Gbps |
| **EU – US East**   | 460 Gbps | 185 Gbps | 980 Gbps  |

For completeness, the tool used to perform the downloads was aria2 and the workload management system was HTCondor. The HTTP protocol was used for all transfers.

## 3  Importing data into Cloud instances

Scientific applications often need significant input data and that input data typically resides on on-prem storage. Understanding how many Cloud instances can be efficiently fed from such storage is thus very important. Assuming that the on-prem storage is capable of serving a large number of clients, the limiting factor is the network throughput between such on-prem storage and the Cloud instances. Two major tests were performed, one against a geographically distributed on-prem storage cluster and one against storage located in a single on-prem location. In both cases, about 40 compute instances were concurrently running, spread among all three major Cloud providers, namely AWS, Azure and GCP, and in regions covering the US, Asia and Europe.

The first test used the distributed nature of the Pacific Research Platform (PRP) [3] as the on-prem storage. Files from the distributed CEPH cluster were exported through Apache httpd on each of the participating nodes, scheduled as pods using Kubernetes. In order to minimize the disruption to the rest of the system, the fetched dataset was quite small, easily fitting into the memory cache on the httpd nodes; this was deemed acceptable, since only the networking between the httpd nodes and the Cloud instances was of interest. The results in Figure 2 were obtained using 14 nodes spread across Southern California and demonstrate that it is possible to reach 100 Gbps using such a setup. The long tail is due both to tests fetching a fixed amount of data and having server nodes with a mix of network interface cards (NIC), including 10 Gbps, 40 Gbps and 100 Gbps.

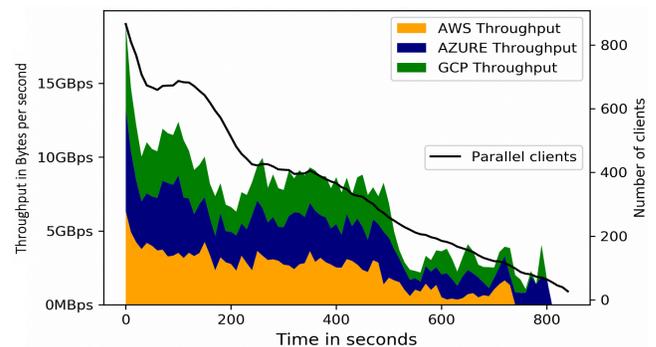

**Figure 2: Observed throughput reading from PRP's distributed storage into instances on the three public Cloud providers.**

The second test was instead aimed at demonstrating that it is possible to fill the network capacity of a single site, exporting from a production storage infrastructure. The on-prem storage for the test was contributed by the Wisconsin IceCube Particle Astrophysics Center (WIPAC) [4] and was located at the University of Wisconsin – Madison (UW), which has 100 Gbps network capacity toward the rest of the world. WIPAC operates a large Lustre pool and we exported a directory tree using 6 nodes with 2x 25 Gbps NICs, on which we ran a load-balanced Apache httpd. This time the fetched dataset was composed of over 80k files and exceeded 1 TB in size, resulting in virtually no memory caching on the httpd servers. The tests showed that it is possible to approach 100 Gbps throughput reading from the Clouds also with such a setup, see Figure 3. The same figure shows that different public Cloud providers have different pairing agreements in place with the University of Wisconsin – Madison.

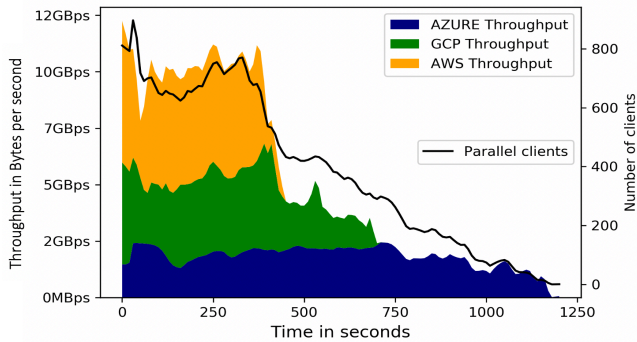

Figure 3: Observed throughput reading from IceCube's UW - Madison storage into compute instances on the three public Cloud providers.

For completeness, the tool used to perform the downloads was aria2 and the workload management system for the Cloud compute was HTCondor. The HTTP protocol was used for all transfers.

## 4 Fetching data from Cloud storage

As described in Section 2, Cloud storage is very performant, and is thus an ideal place to store data. Most science applications however do not live 100% of the time in the Clouds, so having an understanding of how fast data can be exported outside the boundaries of the hosting Cloud provider is very important. Tests pulling from on-prem resources are analyzed in Section 4.1. One may also want to use Cloud storage from a single location to feed compute jobs running in instances located in multiple Cloud regions, even belonging to other Cloud providers. That aspect is analyzed in Section 4.2. Throughput is however not the only factor to consider. Unlike ingress, egress is a metered and billed item in public Cloud realm. This aspect is analyzed is Section 4.3.

All tests used aria2 to perform the downloads and the workload management system was HTCondor. The HTTP protocol was used for all transfers.

### 4.1 Exporting data from Cloud storage

The first set of tests measured the throughput of fetching data from object storage operated by the three public Cloud providers, namely AWS, Azure and GCP. In order to minimize the impact of limited local network throughput, the tests used about 20 pods running on multiple nodes in the PRP Kubernetes cluster, similarly to how multiple Cloud instances were used for tests in the previous sections. All nodes were located in southern California, served by the CENIC network.

Object storage from all three public Cloud providers delivered a throughput of about 100 Gbps. A snapshot from PRP's Grafana monitoring of one of the tests is provided in Figure 4. No significant difference was observed when pulling from the US West and US Central areas, see Table 2.

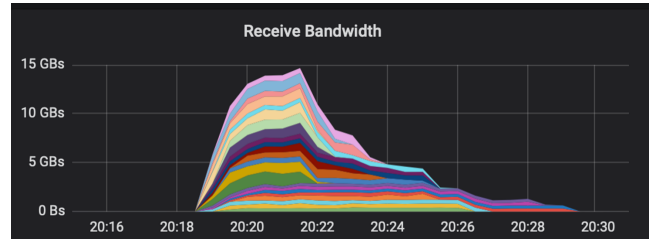

Figure 4: Observed throughput reading from Azure US Central (SouthCentral) Object Storage into pods running on PRP's Kubernetes cluster.

Table 2: Peak Throughput observed reading from Cloud-operated Object Storage from PRP nodes.

|  | AWS | Azure | GCP |
|---|---|---|---|
| US West | 100 Gbps | 120 Gbps | 110 Gbps |
| US Central | 100 Gbps | 120 Gbps | 110 Gbps |

### 4.2 Inter-Cloud data transfers

When computing using Cloud instances, having a single repository of input data is obviously highly desirable, for ease of management. Section 3 provided an overview of what is possible when using on-prem storage, but that's not the only option; a Cloud-operated object storage system in a single Cloud region can feed compute instances in many Cloud regions, operated by many public Cloud providers. Knowing how much throughput can be achieved in such a setup is thus very important.

The tests were performed using the same setup as in Section 3, but pulling data from a single object storage, which of course was operated by a single public Cloud provider and located in a single region. Object storage in the US Central region was used for all tests. The best performance was observed using GCP operated object storage, exceeding 500 Gbps, as seen in Table 3. Interestingly enough, the GCP object storage provided better performance than on-prem storage for compute instances operated by all public Cloud providers, as seen in Figure 5.

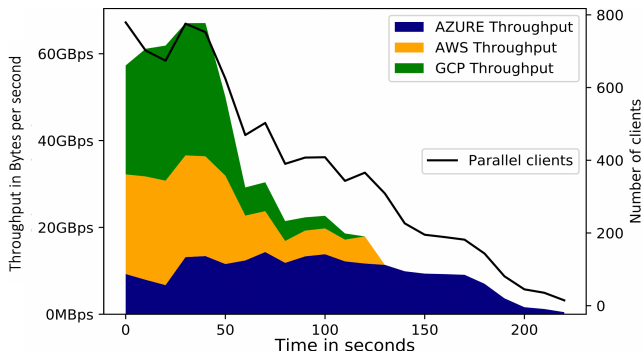

Figure 5: Observed throughput reading from GCP Object Storage into compute instances on the three public Cloud providers.

Table 3: Peak Throughput observed reading concurrently from Cloud-operated Object Storage from compute instances operated by all three public Cloud providers.

| AWS | Azure | GCP |
| --- | --- | --- |
| 320 Gbps | 160 Gbps | 550 Gbps |

### 4.3 The cost of network transfers

All three tested public Cloud providers use a very similar model for networking. To the first approximation, all data transfers inside a single region and all incoming data traffic is free. Tests described in Section 3 thus did not incur any network related cost.

Virtually all traffic leaving a Cloud region is instead billed. The cost changes significantly based on the destination of the data. Moving data to a different region of the same provider is the cheapest, with costs ranging between $20 and $90 per TB. Exporting data to the public internet is significantly more expensive, at $50 and $120 per TB. The inter-Cloud tests in Section 2 and all tests in Section 4 thus incurred a non-negligible network related cost, as seen from the example in Figure 6.

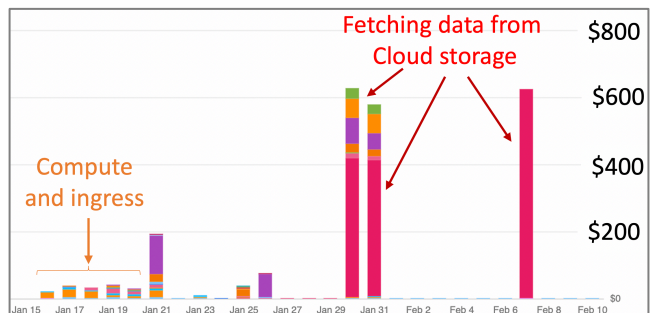

Figure 6: A snapshot from the billing system of one of the public Cloud providers used in the tests.

It should be noted that network related costs can be drastically reduced in a couple of ways. Using pre-arranged dedicated paths, e.g. AWS Direct Connect, can drastically lower the egress costs. Moreover, most public Cloud providers have agreements with most academic institutions to waive network costs if they stay under 15% of the bill, which covers most scientific computing use-cases.

## 5 Conclusions

This paper provides a snapshot in time of what networking of the three Cloud providers, namely AWS, Azure and GCP, is capable of, with an emphasis of moving large amounts of data in and out of the Clouds in parallel. The observed results suggest that 100 Gbps is easily achievable at larger scientific institutions, with measured values available for the Pacific Research Platform and University of Wisconsin – Madison.

When on-prem networking is not fast enough, moving data inside and between Cloud regions provides additional throughput. In some of the tests we exceeded 500 Gbps.

A summary analysis of network related costs is also provided.

## ACKNOWLEDGMENTS

Most of the Cloud costs were covered by credits issued by Amazon, Microsoft and Google.

This work was partially funded by US National Science Foundation (NSF) under grants OAC-1941481, MPS-1148698, OAC-1841530, OPP-1600823, OAC-19044, and OAC-1826967.